# One hundred percent renewable energy generation in 2030 with the lowest cost commercially available power plants

Manfred G. Kratzenberg[1], Hans Helmut Zürn[2], Ricardo Rüther[1]

1 Fotovoltaica UFSC – Solar Energy Research Laboratory, Federal University of Santa Catarina, Florianópolis 88056-000, SC Brazil

2 Power System laboratory UFSC, Federal University of Santa Catarina, Campus Universitário Campus Universitário João David Ferreira Lima – Trindade 88040-900, Florianópolis, SC, Brasil.

We hypothesize that the present expansion of energy generation by variable renewable energy (VRE) power plants, such as wind and photovoltaic power plants, leads to a 100% renewable energy supply in 2030 because of its inherent exponential growth function. This behavior is related to the exponential cost reduction of its generated energy and the nearly unconstrained available potential of its natural resources. The cost reduction results from the continuous improvements in development, research, manufacturing, and installation, also showing a growth of its installation power per power plant or aero generator. We prove that if the historic exponential growth is followed in the future, it is possible to decarbonize the world's electric energy systems' power supply in 2026. Furthermore, the global demand on primary energy can be supplied in 2030, which leads to the total suppression of $CO_2$ emissions related to the energy need of humanity. Because of the related cost reduction, energy costs are not anymore relevant. Our extrapolation is based on the continuation of the historic growth functions of the globally installed PV and the wind power plants, and we also discuss the conditions necessary to enable a transition to such a 100% renewable energy production. Considering a non-constrained growth of VRE power plants' installation power, decarbonization related to energy generation and use can be accomplished in a much shorter time frame as previously scheduled. As a result, climate change mitigation, energy cost reduction, and high employment are attained much earlier than previously planned.

## 1. Introduction

As a result of technological developments in semiconductor manufacturing, related technologies underlay exponential growth principles more than ever. Such a manufacturing results in the exponential increase of the computation capacity based on Moore's law. The related technologies are, e.g., the exponential proliferation of smartphones; the exponential growth of internet users and the exponential growth of the scanning ability of the human genome, in-between others, and they are all based on on the principle the exponential growth of the computational performance

(Diamandis and Kotler, 2012). We think that this behavior of exponential growth is also true for the non-constricted growth of photovoltaic and wind power plants' global installation power, and therefore, its energy generation. As a result, net-zero $CO_2$ emissions can be attained much earlier than expected and scheduled.

Consequently, uncertainties concerning climate change mitigation are reduced, employment is created, and energy is produced at extra low cost as these renewable power plants do have no fuel consumption. Such objectives are essential to hold disaster losses related to climate change at low levels, as discussed in Appendix 1. Furthermore, the expected employment losses because of the impending developments in automatization and artificial intelligence (AI) are compensated. Our proposal results in a rapid transition of the world's electric energy systems based on the growth of the three most important commercially available renewable energy power plants, saying photovoltaic, wind, and hydropower plants. We show that it is possible to generate the world's total demand for electric energy in 2025, solely by these three renewable sources, considering its non-constrained future growth. We also show that the world's primary energy consumption can be generated in 2030, at exceptionally low costs, far below US$10/MWh, by the expansion of such power plants.

Global warming and climate change mitigation are the present most challenging task of humanity. The known relationship to the industrial $CO_2$ emissions, and the unknown outcomes of future climate change, results in potentially existential damage for society (Abdallah et al., 2013). As known, the natural radiative forcing of the Earth's atmosphere provides the Earth's necessary heat energy, which increases its surface temperature for being a habitable planet. However, presently, the Earth's climate system is changing rapidly as a function of extra energy input. The additional energy is a function of radiative forcing, which results in global warming and is a function of the increased $CO_2$ level in the atmosphere. First in-depth research related to this radiative forcing was accomplished by (Hansen et al., 1997), with global simulations of the Earth's thermodynamic system.

The generation of only one MWh of energy emits up to 1000 kg of transparent $CO_2$ into the atmosphere. Even with the current exponential expansion of renewable energy systems, global $CO_2$ emissions are still growing exponentially (Lindsey, 2018). This growth leads to a nearly exponential increase of the $CO_2$ in the atmosphere (NOAA, 2021), resulting in additional radiative forcing, heat accumulation, further temperature increases, which trigger even more significant related climate changes. The resulting radiative forcing increase is currently the most challenging existential threat (Shakoor et al., 2020). Models of related economic losses show in (Stern, 2008) exponential losses of the global domestic product (GDP), which appear as a function of the Earth's surface average temperature increase (Appendix 2). Frist tendencies to the rise of such losses are already verifiable (Smith et al., 2019). Such losses are principally related to climate disasters, and therefore, the emission of $CO_2$ should be reduced as swiftly as possible.

Climate change is one of the most discussed items in the last decades in global scientific and political communities. However, it is still not adequately addressed, as the net-zero CO2 emissions are scheduled only for 2040 or 2055 (Masson-Delmotte et al., 2018). However, it is essential to actualize schedules based on VRE plants' technical and economic state-of-the-art feasibility to mitigate climate change earlier. This feasibility changes rapidly because of the swift development of the related technologies. E.g., the cost of wind and photovoltaic power plants decreased much stronger in the last decades than previously expected (*). Therefore, energy policies should also be actualized for swifter mitigation of climate change, based on the state-of-the-art feasibly for this mitigation.

The concept of providing all energy with hydropower, wind, photovoltaic, solar thermal, and geothermal power plants was first proposed by (Jacobson and Delucchi, 2011) and (Delucchi and Jacobson, 2011), citing further references. The authors' proposal is based on energy storage simulations in a 30 s interval, and they discuss apart from the $CO_2$ benefit also the health benefit because of the reduced city pollution. Their work was followed by the results of (Jacobson, 2017; Jacobson et al., 2015), in which simulations were expanded to several countries, and (Kroposki et al., 2017), which discuss the details as related to the electrical power grid for a 100 % supply of the power grid with renewable and VRE power plants. In their simulations (Jacobson, 2017; Jacobson et al., 2015) use: (i) seasonal heat storage units which provide district heat at costs less than 1 US dollar per kWh; (ii) insulated ice storage units for air condition systems of buildings; (iii) thermally insulated reservoirs of cold and warm water, for air conditioning systems of building complexes such as, e.g., universities; (vi) heat pumps for heating and cooling of rooms, and the production of warm water in residents; (v) demand response to control the heating and cooling of these systems, as well as further non-priority consumer equipment; (vi) hydrogen storage units; (viii) concentrated thermal solar power plants with low-cost rock heat storages, which enable the generation of electrical energy for nearly 24 hours per day; (ix) the storage capacities of pumped hydropower plants; and finally (x) the storage of conventional hydropower plants, where the lower, and the upper limits of its storage lake, as well as, its minimal dispatch to maintain the related river bed, have to be respected. A 100 % renewable power grid should rely, for the most part, on these types of lower-cost storage units, the increase of the consumer's energy price. Each storage type has a range of costs per stored kWh, rather than a fixed cost ((Koohi-Fayegh and Rosen, 2020) – Table 4), depending on the application. Because the lower value of this range is much smaller compared to the cost of lithium ion batteries, we think a large scale use of Litium-ion batteries to provide the necessary storage for a 100% renewable power grid is unrealistic because of the much higher costs this solution presents. This holds true even considering a future exponential cost decrease of lithium ion batteries, which leads to an extrapolated cost of $ 10/kWh in 2030 (Appendix Figure 6). Furthermore, the

large use of Lithium-ion batteries for power grid applications would increase EV costs because of its limited availability.

## 2. Materials and Methods

As principally based on the growth of wind and photovoltaic power plants, in this work, we estimate the time needed for a transition to energy production without the emissions of $CO_2$. We consider (i) power grids supplied by 100% of renewable power plants and (ii) the generation of the world's total energy principally based on VRE power plants considering that its non-constrained present growth patterns can be maintained in the future. Our unconstraint extrapolations are based on the historic exponential growth of the VRE power plants' installation power. We use extrapolated growth figures of the current photovoltaic and wind power installations. We argue that if restraining conditions, such as, e.g., the necessary power lines and the increased generation variability, are accounted for by adequate planning, the future exponential growth of the generated energy by these plants can be maintained. In a visualization with a logarithmic scale, such exponential growth functions lead to a straight line, which we used to extrapolate future installation potential in this graphic. We predict the future generation by these VRE plants by multiplicating the installed power with the average generation capacity factor for each of these technologies. We also propose low-cost proposals for the necessary energy storage to mitigate VRE plants' variability in the power grid based on further references. As a result, this proposal's exponential expansions do not considerably increase the power grid's infrastructure cost. The expansion scenario results in substantial cost reductions for the generated electric energy and the generated primary energy. These low costs permit a present and future exponential expansion of photovoltaic and wind power plants without subsidies.

We believe in the exponential growth rate of VRE power plants, rather than a linear expansion, principally because of the already low cost and de decreasing cost gradients, as inherent to the VRE energy generation. In the next sections, we discuss the factors limiting the exponential growth of VRE power plants and suggest measures to address these limiting factors. We think that these limiting factors do not reduce exponential growth if addressed adequately and corrected.

## 3. Limiting factors that can avoid or reduce exponential growth

As the most assumed limiting factor for high VRE fractions in a power grid are thought the necessary storage systems to operate the electrical power grid, without interruption, because of the variability of its generated power flow. However, low-cost solutions based on state-of-the-art technologies enable VRE support in the power grid showing emulated storage costs as low as only US$ 1/kWh, as discussed

in (Kratzenberg, 2021) section 2.6. Furthermore, high fractions of VRE power plants in the power grid might lead to grid instabilities or generate high costs related to super-peak power plants to compensate for this variability (Hirth et al., 2015). However, as discussed in (Kratzenberg, 2021), several methods keep these instabilities under control. The emulations for a high necessary high short circuit and energy storage capability are most important. The short-circuit capacity can be emulated by exploiting obsolete synchronous generators or the so-called virtual synchronous machines connected to a small battery. The emulation of the synchronous generator's inertia to provide the high necessary short-circuit current as essential for triggering the power grid's protective overcurrent devices enables power grids with high VRE fractions (section 2.7 in (Kratzenberg, 2021)). The availability of resources to expand wind power and photovoltaic power generation is the least problematic because of the many installation locations available, as discussed in (section 2.3 in (Kratzenberg, 2021)). The electric power transmission lines, necessary to be installed to access resource-rich regions for the generation based on wind and solar power plants, are also not limiting if planned adequately with anticipation (section 2.4 in (Kratzenberg, 2021)). The providence of a sufficient workforce for installing and maintaining the VRE plants is also not a growth-limiting factor if educational measures are taken in advance based on the exponential growth (section 2.5 in (Kratzenberg, 2021)). Manufacturing cost is also not a problem as manufacturing does not need to be based on rare-earth elements (section 2.1 in (Kratzenberg, 2021)). Life expectancy is not a problem because of the equipment's typical long lifetime of up to 25 years at the state-of-the-art. Financial constraints to sustain exponential growth are also not a limiting factor because of the secure long-term investments and the already low and falling costs of VRE systems (section 2.8 in (Kratzenberg, 2021)). Political issues are also not a limiting factor for exponential growth because any rational country likes to provide low-cost energy and plenty of workplaces for its citizens (sections 2.9 and 2.10 in (Kratzenberg, 2021)). In a similar form, we think that any reasonable and rational company that is engaged in the energy sector likes to be in business with the VRE technologies because of its substantial price decline, which leads to advantages in the bidding process and the business of power plants (section 2.10 in (Kratzenberg, 2021)).

## 4. Results

In the next sections, we first evaluate the possible energy generation potential of wind and photovoltaic power plants. Then we address the expansion of the cumulative installed power of the essential VRE power plants, which are wind and photovoltaic power plants. Then we compare its present and future energy generation ability based on average capacity factors of the global installation of these power plants. We extrapolate wind and photovoltaic generation individually based on its present exponential growth and combine the extrapolated generation with the extrapolated hydropower generation to access possible future energy generation scenarios obtained based on the extrapolation of its historical expansion curve.

## 4.1 Photovoltaic power plants and systems

The renewable energy generated by photovoltaic power plants is expandable with virtually no limits, which would allow supplying the total energy demand of the human activity's related energy needs in 2030 and many years beyond. E.g., if considered an energy generation solely by utility-scale grid-connected photovoltaic power plants. Based on the exponential growth of these plant's installation power, a total electrical energy that confers to the electrical energy demand in 2032 (Figure 5) could theoretically be generated with installations on a desert surface of 357,667 km$^2$ (Annex 2). Furthermore, the primary energy needed in 2032 could be generated by PV plants on a surface of 2.015 million km$^2$ in occupying less than 6.83% of the world's deserts. Considering, however, the reduction of the primary energy by 42.5% as based on higher energy efficiency (Jacobson et al., 2017), then the occupied desert area would be much lower with 3.93%. Another interesting effect of photovoltaic plants is that its energy generation cost shows a much larger cost reduction gradient in the logarithmic scale as in comparison to wind power plants (Fig. 7b), resulting in a present cost which is similar in comparison to wind power plants (Webb et al., 2020). The much higher decreasing gradient as in comparison to wind power plants (Fig. 8) indicates a much higher expansion rate of this technology. Furthermore, PV plants do not need specific installation locations, as needed in the case of wind power plants, and can virtually added to the power grid in virtually any location. In Figure 1a and 1b we present the historical cumulatively installed power of worldwide installed PV power plants, as based on the same data. While Figure 1a looks as an exponential growing function, Figure 1b shows by the straight line in the logarithmic function space that the installed PV power grows in a truly exponential function over the time axis.

In order to provide a swift decarbonization it is interesting to validate if the historic expansion figures of photovoltaic plants could theoretically supply the world's electricity and primary energy demand, if growing at a similar rate in the future as in the past. The so known Swanson's law predicts that each doubling of the manufactured photovoltaic module capacity leads to a price decline of 20 % (Reichelstein and Sahoo, 2015). As the capacity-price representation in the Swanson's law follow a bi-dimensional exponential function, its presentation in a graph with bi-dimensional logarithmic normalization results also in a straight line, which seems to be a linear function.

However, an exponential PV expansion pattern can also be shown solely as a function of the installation power, where solely the installed power is presented in the logarithmic scale while the time axis is shown in a normal scale (Figure 1a and 1b). If a PV expansion happens at an exponential scale this also results in a straight line (Fig. 1b), which can be easily extrapolated to evaluate the cumulatively installed power of future PV plants.

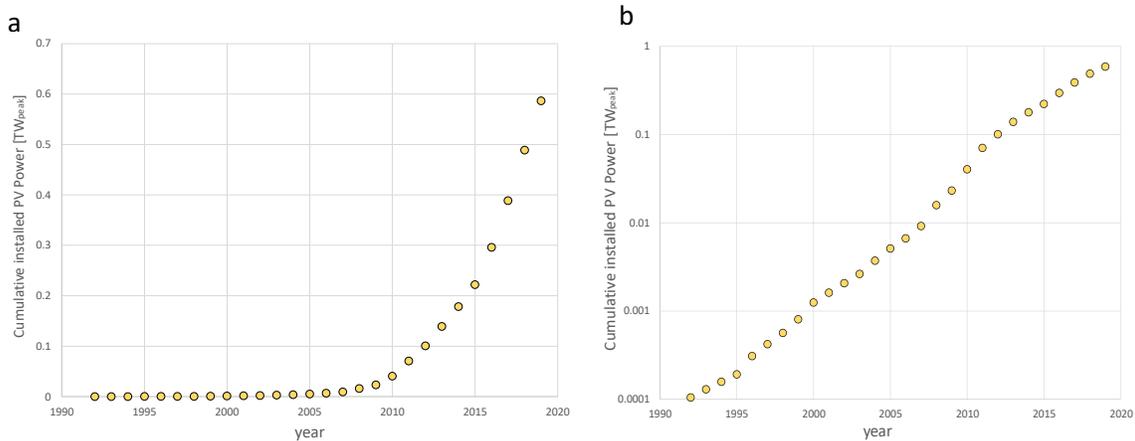

Figure 1: Historic expansion of the installed power of photovoltaic plants in units of Terawatt peak [$TW_{peak}$] as a function of time, as presented in the normal (a), and the logarithmic scale (b). The low deviations to the strait line in the logarithmic presentation proves that an exponential growth is actually happen for the PV power expansion. Data obtained from (Wikipedia, 2021b), citing (Dudley, 2015), (Hashemi and Østergaard, 2016), and (BP, 2020).

To evaluate such an exponential expansion it is, however, necessary to validate if there are any restrictions present that can bend it to lower growth ratios. E.g. for photovoltaic systems such a restriction was present in the past because of the limited available silicon wafers that deaccelerated the PV growth feebly up to 2008 (Fig. 1b). However, this is not anymore the case at the sate-of-the-art as dedicated manufacturing units were implemented to manufacture PV-cell grade specific silicon wafers. Further possible limitations for exponential growth we already discussed in sections 2.1 to 2.10 in (Kratzenberg, 2021) and the discussion shows that in principle no technical, economical, or political restriction can in principle limit an exponential growth of photovoltaic plants, if adequately addressed to.

## 4.2 Wind power plants

In the evaluation of available onshore or offshore potential for the wind power generation, the potential of energy generation ability of each of these potentials is much higher than the expected primary energy consumption in 2030, (Appendix 1). Only onshore wind power plants alone can supply the primary energy demand 3.7 times over in 2030, while the demand on electric energy can be supplied 19.9 times over, which is very promising. Besides the onshore generation, there is a large potential for offshore generation available, which can supply the annual production of 41,200 and 92,500 TWh/year for 20 and 50 m water depths (Arent et al., 2012). The second value can supply the demand on electric energy 2.64 times over in 2030. Using also floating offshore wind power plants, a new modality (Kausche et al., 2018), the offshore potential increases to 192,0000 and 301,085 TWh/year (Appendix 1), while the latter potential is suitable to provide the worlds primary energy 1.6 times over in 2030. In Figures 2 and 3 we show the historic expansion pattern of onshore and offshore wind power plants in normal and exponential scale as a function of time.

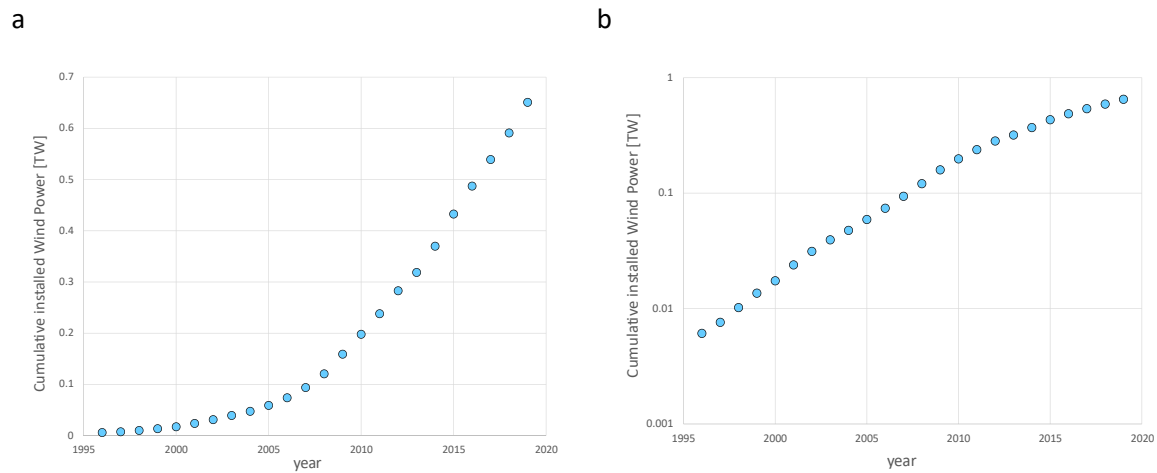

Figure 2: Globally installed wind turbine nominal power in the normal (a) and the logarithmic scale (b), showing an exponential growth up to 2009 (b), which is then followed by a nearly linear growth as to see from (a). As shown by the comparison of the red line in (b), which indicates the exponential growth pattern, in most of the growth history, from 1996 to 2009, an exponential growth was maintained, but starting from 2010 this was not anymore possible to be maintained as probably a result of non-adequate support for exponential growth. Data (black points) obtained from (Wikipedia, 2021e), citing several reports of the Global Wind Energy Council (GWEC).

Wind power plants and photovoltaic power plants present each a cumulative installed power something below 1 TW (Figs. 1 and 2). While the globally installed wind power plants present presently a something higher power as in comparison to the PV power plants, the total of wind power plants installed offshore power is still in its initial growth pattern, showing a much lower installed power. Rather than the total wind power it shows a typical exponential growth function (Fig. 3). By its exponential extrapolation, offshore wind power plants only attain 1 TW of cumulative installed power in 2032. Therefore, they probably do not contribute significantly in a 100% renewably power grid scenario. The low growth function of general Wind power plants installation since 2009 (Figure 2) might be a temporarily slower growth as also happened in the history of photovoltaic plants. If not corrected is can present a limiting factor for exponential growth of VRE plants, principally in this initial stage.

Therefore, the origin of this limited expansion should be identified and corrected as swiftly as possible to return for an exponential growth pattern. One restriction could be the failing skilled labor force (Jagger et al., 2013), but one or more further growth-limiting factors (sections 2.1 to 2.10 in (Kratzenberg, 2021)), e.g., the non-availability of transmission lines, overzealous expansion planning of the power grid, or investment restrains, principally in development countries, could also play an important role. This behavior calls for action for the identification and correction of theses limiting factors.

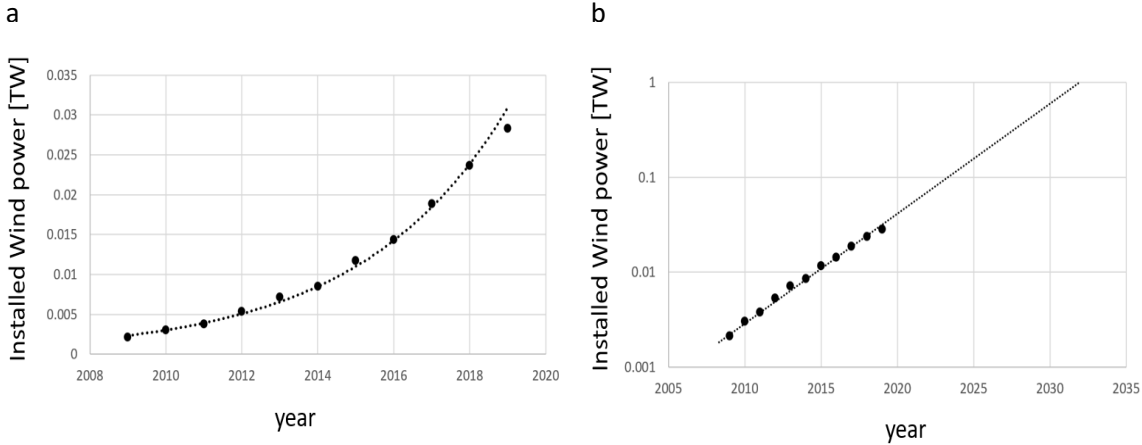

Figure 3: Globally installed offshore wind turbine nominal power in the normal (a) and the logarithmic scale (b), showing an exponential growth pattern. Data (black points) obtained from (Statista, 2021).

## 4.3 Future expansions with VRE power plants

Here we evaluate, if the exponential growth of the commercial available VRE power plants, can much earlier as presently scheduled (i) decarbonize the world's energy generation by the complete substitution of generators and further use of fossil fuel. First we investigate if the exponential growth of photovoltaic or wind power plants alone can supply the world's demand on electrical and primary energy considering its growth based on its historic exponential growth patterns. In Fig. 4a we compare the historic expansion of the installed power of wind, and photovoltaic power plants.

Then by use of the plant's global average capacity factors, we calculated the generation capability of each power plant type and compare these capabilities in Fig. 4b. In Fig. 5 we plot these annual generation capacities in a logarithmic scale, and by exponential extrapolation functions we also show the VRE plant's future annual generation capability in this figure. As hydropower plants grow at a slower pace because of its reduced potential, we used a polynomial expansion for these power plants. The three upper horizontal lines in Fig. 5 indicate the future predicted energy consumption.

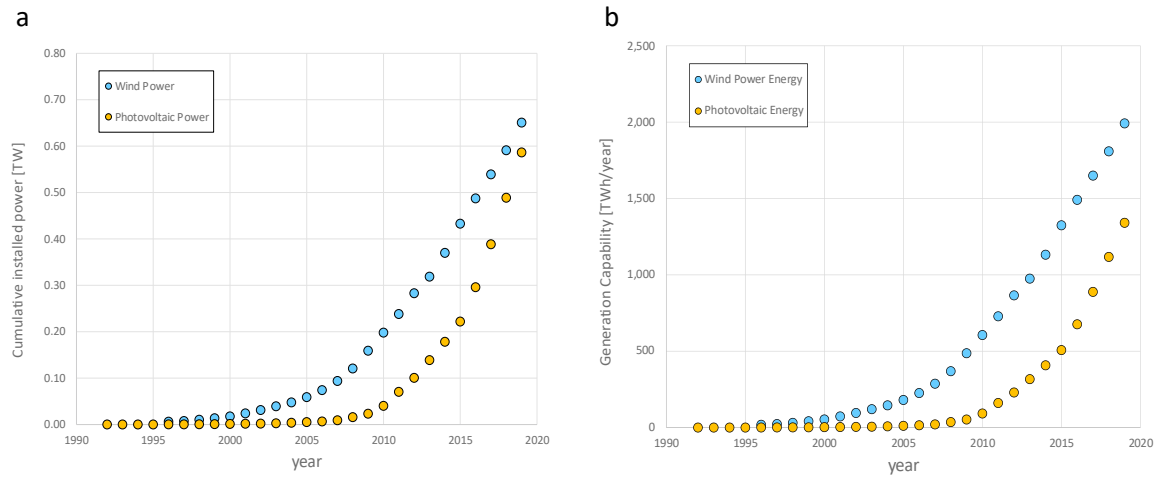

Figure 4: Historic values of the wind (blue), and photovoltaic (yellow) power plants' installed cumulative generation capabilities. The generated energies were calculated with the samples of (a) and the use of average capacity factors of 25.6%, and 35.4%, for the PV, Wind power plants (EIA, 2020).

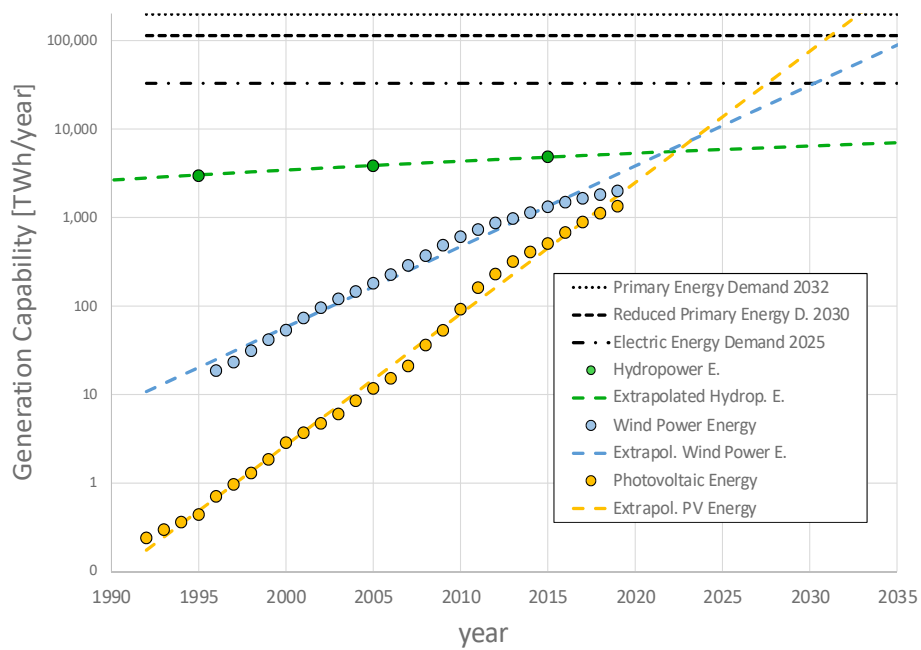

Figure 5: Energy generation capability of the historically installed wind and solar and hydropower power plants, in units of TWh/year, as estimated by using measured average capacity factors, of these technologies. The lower horizontal line indicates the word's estimated consumption of electric energy in 2026 (33,000 TWh/year) and its primary energy consumption in 2032 (198,000 TWh/year). Same generation data as in Fig. 3. Predicted demand on electric and primary energy as presented in (Schalk, 2019). Annual generation of electric energy, estimated by use of the average capacity factors of these technologies and the installed power of these power plants. Same data as in Figs. 1 and 2. Hydropower plant data were obtained from (IHA, 2020), considering a capacity factor of and 43%.

The lower threshold shows the expected world's consumption of electric energy in 2025, and the upper threshold shows the expected total energy consumption, also called primary energy consumption, for the year 2032. The line in-between these two horizontal lines shows a 42.5% reduced primary energy consumption, because of the higher energy efficiency related principally to the fully renewable generation as shown in (Jacobson et al., 2017). When considering a more realistic combined expansion of these most important renewable power plants, electric energy can be already supplied in 2025 solely be the expansion of these power plants (Fig. 6), providing therefore a complete decarbonization of the word's electric energy generation. Furthermore, the combined generation can supply the world's primary energy in 2030 because of the higher efficiency when the primary energy consumption is also based on electric energy.

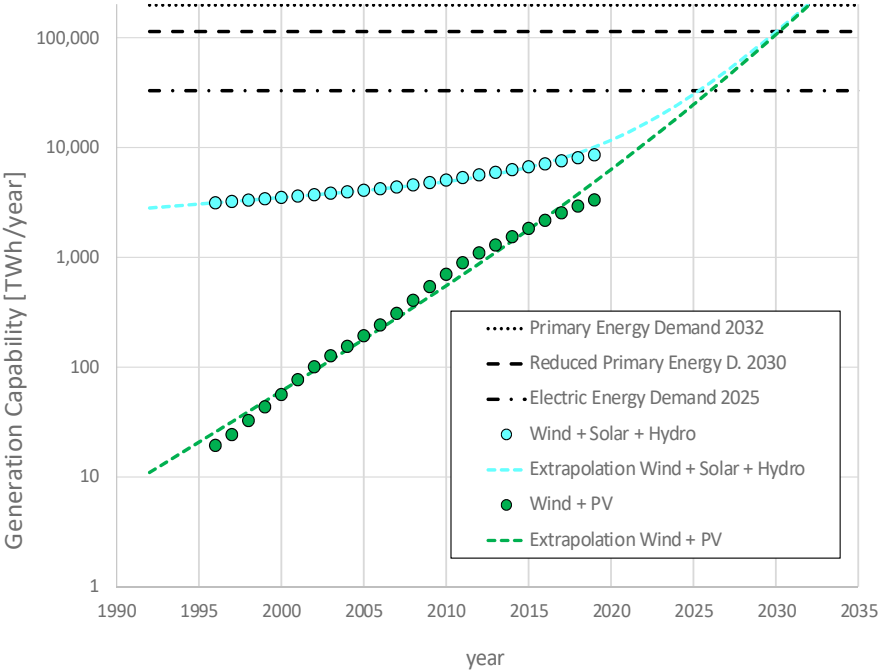

Figure 6: Energy generation capability of the historically installed wind and solar power plants combined, in units of TWh/year, as estimated by using measured average capacity factors, of these technologies. The lower horizontal line indicates the word's estimated consumption of electric energy in 2026 (33,000 TWh/year) and its primary energy consumption in 2032 (198,000 TWh/year). Same generation data as in Fig. 4. Predicted demand on electric and primary energy as in Fig. 5.

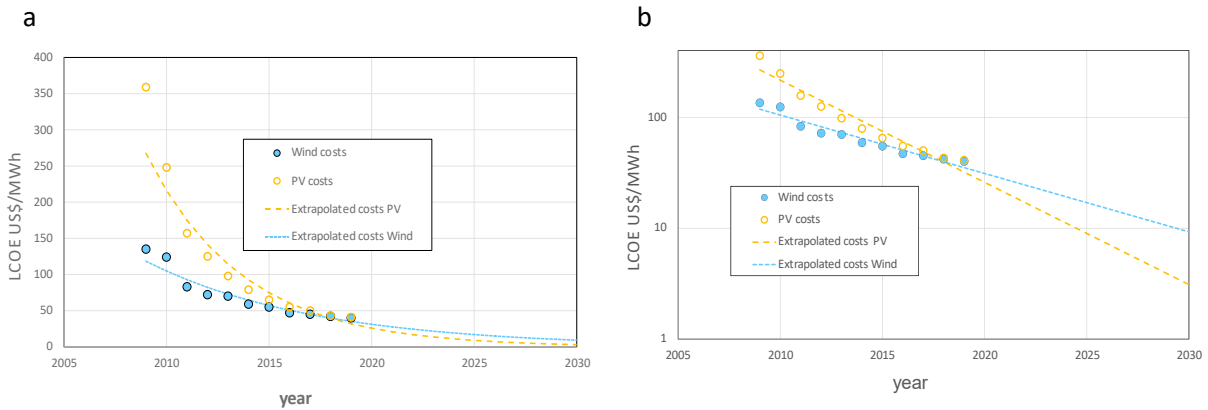

Figure 7. Comparison of the decrease of the Levelized cost of electrical energy (LCOE) of wind power and photovoltaic plants, in the normal (a), and logarithmic scale (b), as a function of time. The dashed exponential fitting lines indicate the exponential cost declines as a function of time for both generators. PV power plants present a higher exponential cost decline than wind power plants. Average LCOE cost samples obtained from (Ray, 2019), citing (LAZARD, 2019).

Fig. 5 also shows that the generation of photovoltaic was insignificant in the past, if compared to wind power systems, presenting more than an order of magnitude lower energy generation at the beginning of its expansion. However, because of the higher growth of its energy generation, in the future the generation by photovoltaic power plants will be most dominant showing quasi an order of magnitude higher energy generation than by wind power plants. Such a dominance can be explained by the higher cost decrease of the photovoltaic power plants if compared to wind power plants, as shown in Fig. 7. The interpolation lines show that wind and solar power plants show a decrease of its LCOE at an exponential scale. However, PV power plants show a much higher gradient of this exponential decrease. Additionally, the strong curve behavior principally the PV plants show a more than exponential decrease of the energy generated by PV power plants (Fig. 5b). A similar cost behavior can be read from Fig. 8 in a combination with the PV plant expansion. The so called Swanson's learning curve, shows the PV module costs as a function of the PV module shipping in a bi-logarithmic curve representation (Diamandis and Kotler, 2012). Fig. 8 is similar but instead of the PV module cost it presents the unsubsidized Levelized cost of electric energy (LCOE), for Photovoltaic systems, which also includes the further system costs such as inverter, mounting rack, land, installation, and maintenance cost.

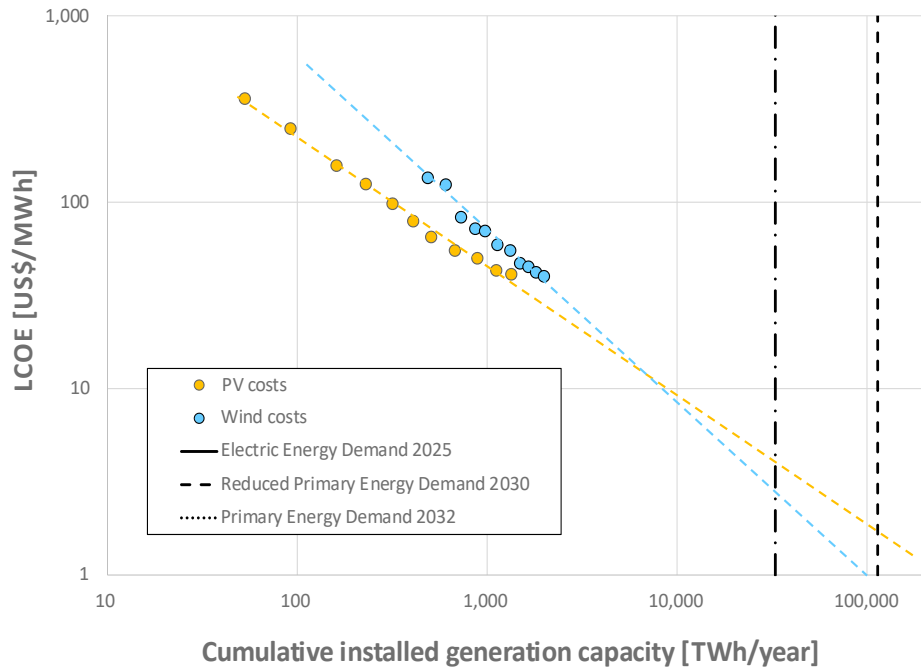

Figure 8: Swanson's learning curve, here expressing the Levelized of electric energy (LCOE) as a function of the generation capacity of the cumulative installed capacity of (i) photovoltaic plants (orange), and (ii) wind power plants (blue), in 2009 to 2019, and its extrapolation curve in the bi-logarithmic scale. The vertical lines on the left side indicate threshold values of the total electric energy consumption as predicted for 2026 and the word's primary energy consumption predicted for 2035.

Furthermore, on the abscissa we present generation capacity of the cumulative installed PV systems in units of [TWh/year], instead of the cumulative shipped PV module power. Such a presentation gives a clearer view of the end product saying the generated energy and it also allows a comparison of development of further technologies such as the wind power also shown in Fig. 8. The figure shows that as a function of the installed generation capacity, PV systems present in the whole diagram a lower cost as in comparison to wind power plants. This means the historically higher costs of PV modules (Fig.7a), where only present because of its lower generation capacity, if comparted to wind power plants at that time. However, based on the crossing of the two lines in Fig. 8 we expect that wind power plants generate energy at lower cost than PV power plants in the future. Therefore, it is important that efforts are taken in the near future to grow the presently nearly linear wind power expansion back to the typical exponential growth, enjoying the benefits of a non-restricted exponential growth and its related exponential cost reduction.

## 5. Discussion

We like to remark that the here presented results are based on the extrapolation of the historical installation capacities of wind and photovoltaic power plants. We draw possible expansion scenarios for the near future as based on these historical exponential expansion schemes for a rapid expansion in the near future that are in the range of technical and economical possibilities. However, extrapolations bear uncertainties and, therefore, the future installed cumulative capacities of VRE power plants is function of the worldwide efforts taken to provide its unlimited exponential growth.

### 5.1: Photovoltaic generation versus combined wind and PV generation

We showed that the non-constraint growth of wind, PV, and hydropower plants leads to net-zero of the electric energy and primary energy generation by 2025 and 2030. From Fig. 5 can be seen that the generation capability of PV plants initiated its growth at a much lower capacity if compared to hydropower and wind power plants. However, as the PV capacity grows at a much higher rate in comparison to wind and hydropower plants (Fig. 5), it will play the most important role in the future. The second largest VRE energy will be generated by wind power plants. Actually comparing PV and wind plants within the same time interval from 2009 to 2019, it can be seen that the wind power plants generation capacity has grown approximately one order of magnitude, while the PV capacity has grown up to two order of magnitude (Fig. 8).

In our energy estimation of the generation capacity of the different power generator technologies, we used the cumulative installed power of a technology and multiplied this with the global average capacity factor. However, exact capacity factors depend on the installation location, the used system, and its appropriate maintenance. Therefore using a global average capacity factor leads to some small uncertainties. However, the use of the global average capacity factors is a good estimation as based on the total of the presently installed wind, photovoltaic, or hydropower plants, as the global average capacity factors are estimated on the base of a plant type global energy generation. Mainly negative bias uncertainties also exist in the predicted exponential growth character as based on the discussed limiting conditions. E. g., if the discussed growth constraining conditions of sections 2.8 to 2.10 in (Kratzenberg, 2021) are not adequately validated and addressed to, for its elimination in different countries, then the expansion of the renewable energy generation can appear at a much lower growth rate as here predicted. Limiting conditions for less than exponential growth already appeared for wind power plants within the last 4 years (Fig. 5), and it is presently unclear if this is solely a temporarily setback, as already appeared for PV plants between 2004 and 2009 (Fig. 5), or is a new tendency. However, the historic growth of PV plants is still very promising, showing even higher than exponential growth in 2008 to 2012 (Fig.5). Therefore, we think that this technology will play the most important role in the decarbonization of power grids and in the generation of the world's primary energy. However, because of the typical knee curve as characteristic for exponential growth we expect in the near future also setbacks as based on eventual constraining conditions for PV power plants. This setback behavior demand for activities in energy policies, regulations, and in the further development of the VRE power plants and its integration in the power grid.

## 5.2 Cost and growth of PV and wind power plants

As to see from Fig. 4, the initial growth of the wind power generation capacity is much higher in comparison to PV generation, but it slowed down since 2017 (Fig. 5), which should be a temporarily growth limitation, because of some presently acting restraining conditions. E.g., the lower generation of PV plants can be explained as in the 1990'er the energy generated by PV plants had a much higher cost in comparison to wind power plants (Fig. 7), being still be present in 2009 (Fig. 8). Only currently the LCOE of PV plants is very similar to that of wind plants (Figs. 7 and 8). We have to note that the exponential reduction of the LCOE of wind and PV plants is a clear function of its installed generation capacity. This is shown by the strong point correlation to the regression lines in (Fig. 8). This means that if a non-constrained growth of these VRE technologies can be guaranteed by international efforts, then a very large cost decline will be the result, providing in 2030 energy by these technologies at something above of 1 US$ / MWh, saying at nearly zero costs (Figs. 7 and 8).

The historical higher generation cost of PV plants in comparison to wind power plants is a result of the much lower installed generation capacity of PV power plants in comparison to wind power plants (Figs. 5 and 8). However the generation capacity of PV plants has grown at a much higher gradient than wind power plants, which also lead to a much higher decline of its generation costs. We think that the higher growth and therefore cost decline of PV plants comes available as PV plants can virtually be installed everywhere and at any size, which is not the case for wind power plants. Furthermore, PV plants are easier to be installed and maintained because of its lower high and as most PV plants have no or little moving components. Comparing the plant costs independently of time but as a function of the installed generation capacity, the energy generated by PV power plants presented historically always a lower cost as in comparison to wind power plants, which is a new insight that can only be discovered by the proposed comparison of Fig. 8. The figure shows also that the PV technology already arrived at the cost of wind power plants even with a lower installed generation capacity. This means the PV technology is actually the lower cost technology on a global scale and the cost of PV plants was only higher, starting from the 1990er, because of the much lower installed generation capacity. The today's similar energy generation cost of PV plants in comparison to wind power plants is principally a function of the higher cost decline of PV in comparison to wind power plants (Figs. 7 and 8). However, if wind power plants grow exponentially then in the future very similar generation costs are expected, and wind power plants can probably generate energy at a something lower cost (Figs. 7 and 8).

## 5.3 Exponential growth and lags in exponential growth

Presently wind power plants show a large time lag in its exponential growth of approximately 3 years (Fig. 5). We think that this time lag appeared because of a constraint growth behavior, as function of one or more of our listed restraining conditions discussed in (Kratzenberg, 2021). E.g., the failing of sufficient energy auctions that allow wind power projects to be installed. If these conditions are adequately accounted to and corrected, we think that this grow function can be corrected and its grow function be bend back to exponential growth, similarly as the PV capacity was bend back after solving the restricting material supply. Differently, photovoltaic systems can be installed without auctions in many residences, which might be one of the reasons than there are no time lags presently in the growth of the cumulatively installed PV power.

## 5.4 Single and combined expansions of VRE plants

Theoretically the world's demand on electrical and primary energy could be generated solely by PV power plants in 2027 and the midst of 2036 (Fig. 5). However, the most probable and useful scenario is a combined growth of wind, photovoltaic and hydropower plants, as well as the growth of further renewable power generators. Its complementary energy generation profile is especially useful to reduce the demand on storage capacities because of the plants seasonal and spatial generation complementary. Such an expansion figure considers, however, that the tendency of linear growth within the last four years of the wind power installations (Fig. 5) can be bend back to exponential growth by adequate research, coordination and management efforts of the global, and country-wise, VRE expansion.

## 5.5 Outcome of non-constrained growth of renewable power plants

A higher than exponential growth of offshore wind power plants could be a solution to correct the present setback of the growth curve of the wind power generation. This would need, however, government incentives for very large scale manufacturing of the infrastructure necessary to install offshore wind power plants. Therefore, much easier to accomplish is a non-constrained onshore generation capacity growth, because the necessary infrastructure for the wind power installation is also used for a series of further construction activities. However, the offshore wind turbines have typically a larger size, which results to higher capacity factors, which facilitates the offshore growth. We think that onshore size constraints can be overcome in exponential growth scenario by near-power-plant manufacturing-units combined with short distance air-transport of rotor blades.

Meanwhile, we expect that PV power plants will generate more energy than wind power plants starting from 2024 as to see by the crossing of the straight lines of the extrapolation curves of wind and PV plants in Fig. 5. Therefore, a lower than exponential growth of the wind power plants would only present a minor setback. Meanwhile, a combined exponential growth of wind and PV plants alone could theoretically supply the electric energy demand in 2026 (Figure 6). Accounting also the expected hydropower plants growth, these three plant types are able to supply the word's electric energy need in 2025 as to see from this figure. Considering a further unconstrained growth of these renewable power plants, such an expansion leads to a fully renewable primary energy matrix in 2030. In such a scenario the inherent nearly zero cost of electrical energy (Figs 7 and 8) is used for industrial, commercial, and residential heating and cooling; for any type of transport and for any further energy use globally. If further renewable power sources are also accounted to such a feat is accomplishable in some months earlier because of their much lower historical expansion.

The continuous growth of the rotor blade lengths of wind turbines, not only to increase its installation power per unit, but also to explore lower wind velocities for offshore wind power plants. However, the power per unit of onshore wind power plants is presently stagnant because as very long blades cannot be transported, because of traffic infrastructure limitations. As the power of the wind turbine installations grows exponentially, very large wind turbines should ideally also be installed onshore. Therefore, in exponential growth scenarios we think that it will be helpful to manufacturer not only the wind turbine towers (REVE, 2020), but also rotor blades in locations close to the areas of high wind energy potential. In such a case we think that short distance air transport of rotor blades can become economically available. The air transport's higher costs will be less significant because of the short distance and the strong cost decline of the generated energy in the mass production scenario. Furthermore, the cost of the air transport with helicopters will be outweighed by the higher efficiency and capacity factors of wind turbines with longer blades that allows the exploitation of lower wind speeds. Such transport issues are no present for large and very large photovoltaic power plants, as pre-assembled mounting racks do not imply any space limitation for its transport, because of its smaller size and its modularity.

## 6. Conclusions

We have shown that the extrapolation of the present exponential growth pattern, of wind and photovoltaic power plant's power and generation capacity can supply the global demand on electric energy in 2025, and primary energy in 2030. As a result we have shown a net-zero $CO_2$ emission scheme, as being an important part of the mitigation of climate change, in a much shorter time frame as presently considered in 2050, e.g. in (Huang and Erb), (Kusch-Brandt, 2019). This scheme is on the main part based on the energy generated by VRE power plants, combined with extra low-cost storage units. Because of the already low and the exponentially falling cost per MWh, we think that the cumulative installation of PV and wind power plants may even grow higher that on exponential behavior if, however, the constraint's limiting factors as discussed in (Kratzenberg, 2021) are adequately addressed to. We show the effectiveness of wind and photovoltaic power plant installation, in visualizing that solely the wind or the PV plant's growth alone is able to supply the world's electric and primary energy demand. Such a configuration would demand however at least 3 years more for a 100% renewable supply mainly by PV and to a lower extent by hydro and wind power plants (Fig. 5). However, such a case would result in higher costs for a 100% renewable energy supply because of the higher demand on storage capacities to store energy at night and the presence of solely the spatially complementary generation of PV plants, while the generator complementary is not explored in this case. Furthermore, for very large shipping of wind power plants its cost per generated MWh is expected to decrease below that of PV plants (Fig. 8). Therefore, the exponential growth of wind power is important and the present time lag in its exponential growth must be evaluated more thoroughly and corrected, based on the discussed limiting criteria.

We think that if the constraints related to this criteria are eliminated adequately, its present linear expansion is only of temporal character and exponential growth curve of the cumulative installed power and generation capacity. The most probable case for the future is the exponential growth by photovoltaic and wind power plants, which are able to provide a complete reduction of the $CO_2$ emissions as related to the energy generation. We estimate that if the world's primary energy can be supplied by the presented scheme in 2030, the global $CO_2$ emissions would be reduced by approximately 73 %, as the present energy generation is responsible for 73.2% (Ghosh, 2020) of the carbon dioxide emissions. Therefore, a primary energy generation without $CO_2$ emission can significantly accelerate the mitigation of climate change. Combined with adequate reforestation measures for $CO_2$ capture (Fawzy et al., 2020), $CO_2$ emission could eventually be reduced to preindustrial scale. This results to (i) the mitigation of climate change and its related risks, (ii) generate additional income, from the generated low cost energy and wood, and (iii) provide additional workplaces.

Principally because of (i) the present exponential growth of VRE plants, (ii) the present and especially future economic figures, and (iii) the available resources, saying generation potential, we think that its exponential growth is very attractive and, therefore, not stoppable in the future. Such a behavior will be of large benefits for a sustainable economic growth and the upscaling of employment. In the recent past such an exponential growth is especially shown for PV and offshore wind power plants. However, non-adequate energy policies, non-tuned regulations, and non-adapted power grids could limit such an expansion, principally for high VRE generation fractions. Therefore, we think that it is important that the international society envisions favorable and normalized energy policies, technical standards, and regulations that enable power grids to integrate large generation fractions of VRE power plants. Furthermore, international technological and research cooperation's should be established for the exchange of experiences and technology, especially for the power grid adaption. As a result of these measures we think that exponential growth restrictions for the expansion of wind and photovoltaic power plants can be avoided. Adequately high carbon taxes that are appropriate for the compensation of the present and future economic and environmental damage should also be considered. As discussed in (Hansen et al., 2015), an internationally homologized carbon tax should be established, and carbon taxed should also added to the import of products in a proportion as weighed by the product price, the country's emitted $CO_2$ and its gross national product. Additionally, subsidies of thermal power plants, and principally coal fueled thermal power plants, which produce the largest $CO_2$ per MWh generated, should be internationally abolished. The long term generation commitments related these power plants should also be avoided, inclusively to avoid economic losses.

Our exponential grow curves and the discussed growth constraints of the analyzed VRE technologies (Kratzenberg, 2021) can be used as a guideline to validate if there are limiting factors that reduce the exponential growth of VRE technologies in the different countries. Furthermore, we expect that the nearly linear growth of the wind power expansion in the past three years, can be bound back to its exponential growth pattern if the essential requirement of adequate energy policies are established. Such polices should act on the control and elimination of the discussed limiting factors for exponential growth. The possible reasons of the bow from originally exponential growth, back to linear growth in the last three years, should be investigated in different countries and positive experiences should be communicated and applied in an effective way allowing exponential growth in most of the countries. We consider these measures as being effective for a rebound to the original exponential growth pattern of the cumulative installation of the global wind generator power. Together with the exponential growth of PV this results in the significant reduction of future $CO_2$ emissions, and its related radiative forcing effect in the atmosphere, which enables the mitigation of severe outcomes of future climate change (Appendix 2).

The generation cost (LCOE) of wind and photovoltaic systems will decrease at an exponential scale furthermore, which is shown best by the clear cost decrease tendency as a function of its manufacturing scale in Fig. 8. It is also shown by the Swanson law, e.g., in (Diamandis and Kotler, 2012), showing an exponential price decrease of the PV module cost as a function of the cumulative installed capacity. However, Fig. 8 presents apparently a higher point correlation if comparted to the

Swanson's curve. The figure also shows that for exceptionally large generation capacities the generation cost of wind and photovoltaic energy have with the expected ~ 1 US$/MWh virtually near zero costs. In such scenarios the storage cost, as well as the maintenance and control of the power grid will represent the main part of the future electric energy costs. However, in the case of the adoption of a large quantity of heat storages controlled by DR (section 2.6 in (Kratzenberg, 2021)), which emulates an electrical storage, the cost of the electric energy for the final consumers will decrease significantly, because of the ultra-low cost of storage and generation of energy. This low cost structure also enables the economic generation of hydrogen and methane not only to store excess electrical energy, but also for its use in long distance transport, and for further industrial processes. Additionally, such low costs allows the use of the excess energy in large desalinization plants, providing additional drinking water and water for irrigation in the agriculture (Diamandis and Kotler, 2012), and allows a further cost decrease because of the even higher mass production of PV and wind power plants. From the extrapolation of the recent growth pattern it can be seen that the expansion of photovoltaic power plants, will be able to provide the main part of the world's primary energy in 2030 (Fig. 5). Therefore, even if wind power plants do not grow on an exponential rate, a 100% primary energy generation can be attained in-between 2030 (Fig. 6) and 2032 (Fig. 5). However, the growth of wind power plants is important because of its complementary generation in different time scales, which lowers the necessary storage capacity and its cost. Such a relationship is easy to see, as without wind generation a large part of PV generation would be necessary to be stored for its supply over-night and in periods with low solar irradiance.

The storage of conventional and pumped hydropower plants will also play an important role in a 100% renewable power supply. However, because of the limited expansion potential of the global hydropower plants (Appendix 1) off-river hydropower plants will be especially important in countries, which current energy generation is mainly based on non-renewable power plants. Meanwhile, to minimize the costs of the necessary storage, the much lower-cost thermal storages units, as emulated by DR, should be implemented by market forces on the base of subsidies, as this configuration will finally result in a much lower cost of the electric energy for the energy consumers. Additionally, power girds should be enlarged and interconnected to make use of different consumption profiles and generation profiles of the VRE plants that also reduce the storage demand.

The most important use of lithium ion batteries is in Electric Vehicles and to attend the demand on electric energy at charging stations in parking lots, at locations where the power grid is not already actualized to accomplish this job, or in locations still without power grid, where the battery is charged with a renewable power plant. Such a concept not only enables the rapid implementation of an EV infrastructure, but also results in the real demand on the necessary power grid infrastructure, allowing the use of battery containers in several different parking lots within its lifetime, which results in higher ROI (ROI). Batteries can also be important to provide auxiliary functions to improve the power grid's quality via specific control functions (do Nascimento and Rüther, 2020) or virtual synchronous machines.

Energy security was an argument in 2020 (Pinto, 2020) to mitigate possible problem of a brownout in the Brazilian electricity grid by the installation of thermal power plant. A brownout is a blackout by reason of insufficient generation capacity rather than by reason of technical problem, such as a transmission supply breakage. However, in the present case the energy generation cost of PV power plants is already below the cost of thermal power plants. Furthermore, a correctly tuned power grid generates more energy in dry periods, by the installed PV and wind power plants, as in periods with average hydropower generation. This excess generation leads to lower use of the hydropower generation capacity conserving therefore high water levels in the hydropower lakes. In countries with low generation capacity dry periods are also not a problem, because the used off-river-hydropower plants are only used for the circulation of water between two lakes of different heights, and its evaporation can be reduced by a floating covers in extreme cases, or even by floating PV power plants.

Finally we like to remark than it is important that governments, industry and the area of power grid research take actions to allow a non-restricted exponential growth of the energy generation with VRE power plants. Otherwise, the combined benefit of a future with ultra-low-cost-energy generation, the providence of labor places, and the mitigation of climate change cannot be exploited in its full magnitude.

## Appendix 1 - Basic data for the potential estimation

**Energy demand by 2030:**

- Electric energy demand in 2030: 35,000 TWh/year (Schalk, 2019),
- Primary energy demand in 2030: 186,000 TWh/year (Schalk, 2019).
- Reduced primary energy demand in 2030 because of a 42.5% reduced energy consumption by the 100% renewable energy generation (Jacobson et al., 2017): 106,950 TWh/year.

**Example 1: 2030 Energy demand solely covered by PV power plants:**

- Global desert area, excluding Antarctica, as calculated from (Wikipedia, 2021c), citing several further references: 34.93 million km$^2$,
- 42.8 MW$_{peak}$/km$^2$ installed power on average, as calculated from data of several utility PV plants presented in (Wikipedia), citing several further references,
- Average capacity factor of PV plants in the US is 25.6%, as based on 2017 data (EIA, 2020),
- PV plant area 357,667 km$^2$ to cover electric energy demand in 2030,
- PV plant area 2.015 million km$^2$ to cover primary energy demand in 2030,
- Occupied global desert area 1.21% to cover electric energy demand in 2030,
- Occupied global desert area 6.83% to cover primary energy demand in 2032,
- Occupied global desert area 3.93% to cover reduced primary energy demand.

**Example 2: 2030 Energy demand solely covered by wind power plants:**

- Onshore wind power potential: 690,000 TWh/year (Lu et al., 2009),
- Offshore wind power potential < 200 m: 192,000 TWh/year (floating wind turbines); (Arent et al., 2012),
- Offshore wind power potential < 1000 m (Kausche et al., 2018): 301,085 TWh/year (as extrapolated for floating wind turbines at this depth, Appendix Figure 1),
- Onshore and offshore total potential 301,775 TWh/year,
- Global wind potential necessary to cover electric energy demand in 2030:11.59 %,
- Global wind potential necessary to cover primary energy demand in 2030: 61.63 %,
- Global wind potential to cover reduced primary energy demand: 35.44%.

Presently hydropower as a renewable energy source, has already developed generation potential of 6.5 TWh/year, which is more than half of the exploitable generation potential of 10.5 TWh/year (Mariusson and Thorsteinsson, 1997).

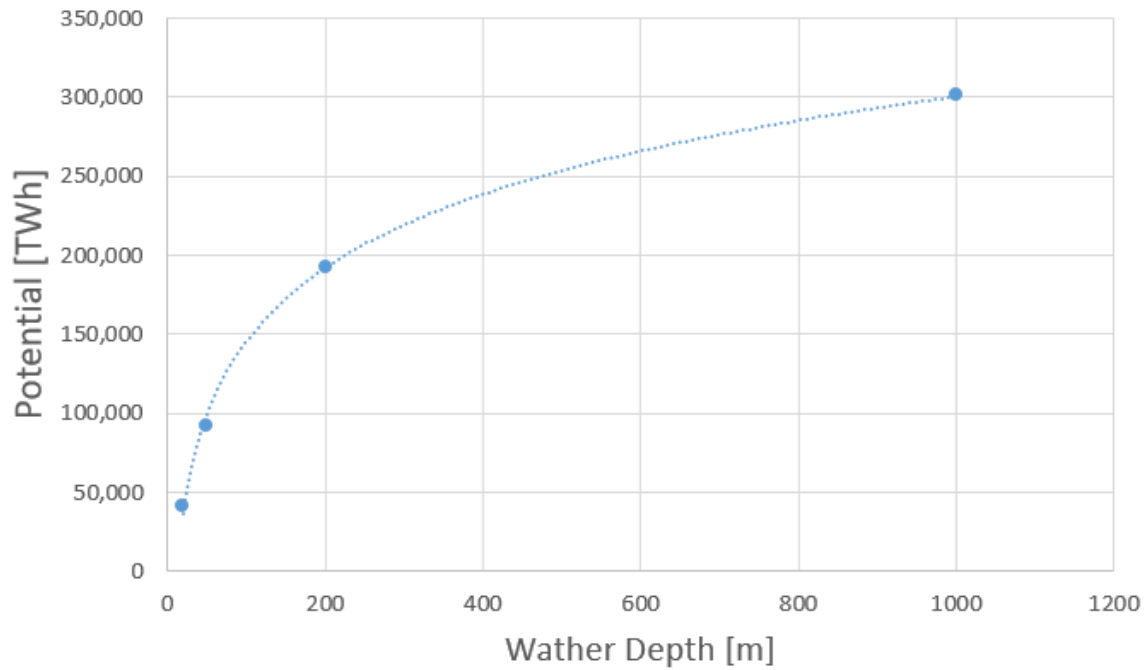

Appendix Figure 1: Total offshore potential considering fixed installation and floating wind turbines. Potential was correlated to the available space for different depths and the value for 1000 m depth were extrapolated with the available sea space for this depth. Data obtained from (Arent et al., 2012) and (Kausche et al., 2018). Floating turbines are considered for the water deepths between 90 and 1000 m, presenting approximately half of the available potential or resource.

## Appendix 2 - Worst case outcomes of climate change

The safety of potentially dangerous technological machinery projected by humans, e.g., commercial aero plans or nuclear power stations, has to be validated by the evaluation of worst case scenarios. However, the future of the Earth's climate is not anymore determined by nature but is rather dependent on the appropriate action of humans. Therefore, worst case scenarios must also be evaluated for climate change.

Presently two thirds of the global warming is a function of the atmosphere's water vapor. However there is a tendency to increase the water vapor in the atmosphere, because of the global warming. The present temperature increase as a function of global warming is still moderate in comparison to what can be expected it the future, as presently the added heat energy is absorbed by the melting of the global glaciers. After the melting of these glaciers heat cannot be anymore absorbed in the same proportion, which leads to a much severe heating of the global atmosphere and more water vapor. Therefore, a possible worst-case scenario outcome is the evaporation of the Earth's available seawater, an effect well known in planetary science, because of the tipping point at which a global warming results in extremely high concentration of water vapor in the atmosphere. This tipping point is also based on a positive feedback, which increases progressively radiative forcing as a function of the increased level of water vapor. It is an outcome to which other planet systems were already subjected to. E.g., as shown by numerical long-term simulation of its planetary circulation system, Venus were habitable some billion years ago and presented an ocean which was, however, evaporated because of the radiative forcing (Wheeling, 2020). Therefore, a conservative evaluation of the long-term outcome of climate change should consider this worst case hypothesis.

Such a worst-case outcome accompanies a sea-level rise up to 70 m, as related to the melting of all the available glaciers on the Earth's surface, meaning the north- south- and mountain glaciers. Without the decarbonization of human activities the $CO_2$ level will exceed the levels of the Ecocene epoch, 56 to 34 million years ago, which can by theory result in a see level rise from 20 to 70 m, probably within the next 2000 years, depending on the quantity of the human-induced greenhouse gas emissions (Le Cozannet et al., 2018).

Another worst case scenario is that of severe economic losses as a function of the average temperature increase of the atmosphere's surface temperature. As based on the indirect measurements of the climate's history by the extraction of the south-pole ice core, (Appendix Figures 2 and 3), the global warming of the Earth's atmosphere at its surface will be nearly $\Delta T \approx 16$ °C under thermal equilibrium conditions (Appendix Figure 3). This equilibrium is solely expected for the extrapolated $CO_2$ level of 2025, and as the present exponential increase of $CO_2$ emission continues, more heating can be expected. For the $CO_2$ level of 425 ppm in 2025 such losses present an estimated value of 70 % of the global gross domestic product GDP as based on climatic events, considering the thermal equilibrium condition of the atmospheres and hydrosphere in any time after 2025 (Appendix Figure 4). First signs of such an exponential increase of these losses are already present (Appendix Figure 5), and its expected exponential increase leads to the almost complete deterioration of the global economic system, surly related to intense suffering of humanity.

Presently the cost of severe storms sums up to a total of US$ 1.875 trillion within the last 40 years, solely the US (Smith et al., 2019). Starting from 2007 this cost is continuously on high values (Appendix Figure 5), which are approximately an order of magnitude larger compared to the interval between 1980 and 2007 (Appendix Figure 5), indicating a dangerous exponential increase of the cost related to future climate events.

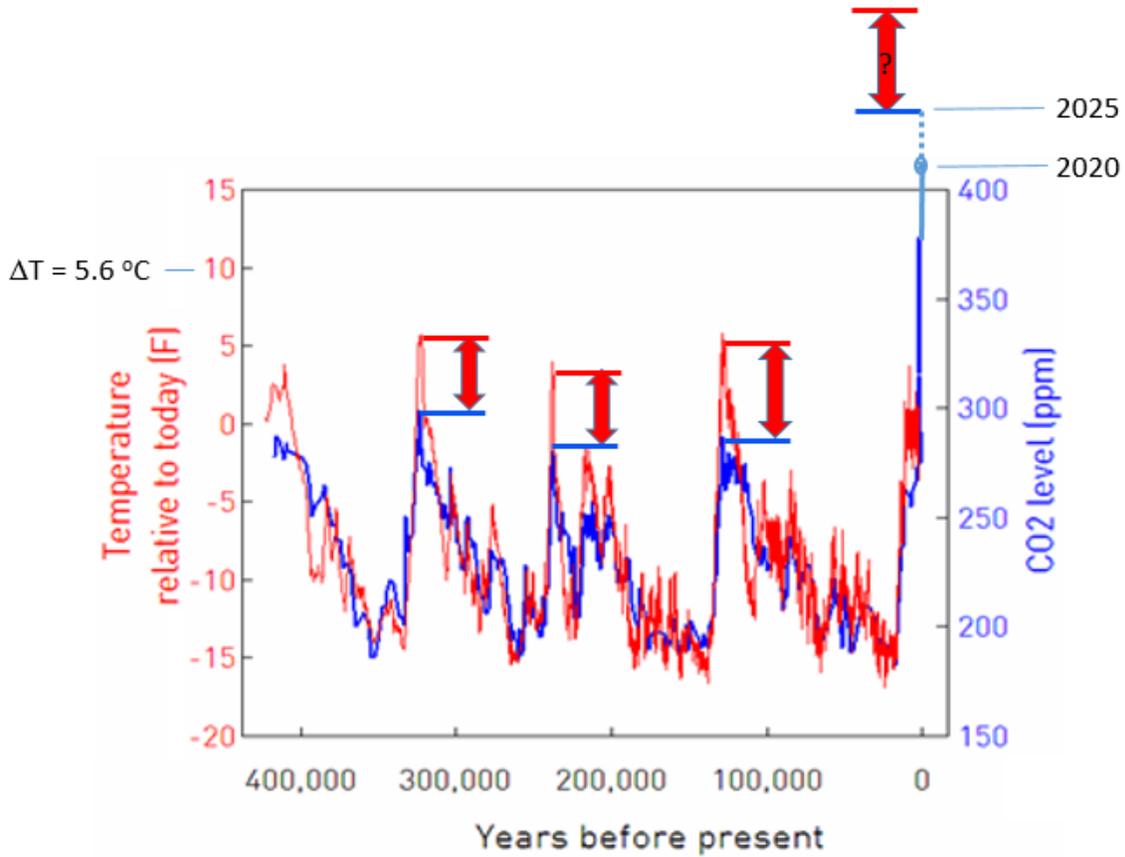

Appendix Figure 2: Superposition in the behavior of the atmosphere's $CO_2$ level and its global average temperature, showing that these variables are highly correlated in the Earth's warm an ice-age cycles. The arrows indicate that the warm cycles present short-term high temperatures with higher amplitudes than the atmosphere's $CO_2$ content. However, because of the present high $CO_2$ level, as grown exponentially since the industrial revolution, there is a large expected temperature response of the surface temperature for the present $CO_2$ level. Solely in 2025, where the expected $CO_2$ level is 425 ppm, the correlated temperature increase is in the order of roughly $\Delta T = 10$ °C, considering the historical pattern of three warm cycles is followed. The measurements are based on the analyses of the conserved layers of the Antarctic ice core, presenting air samples and the heavy water isotope, which increases in global warming periods (Petit et al., 1999).

Source: https://stevengoddard.files.wordpress.com/2011/05/last_400000_years.png

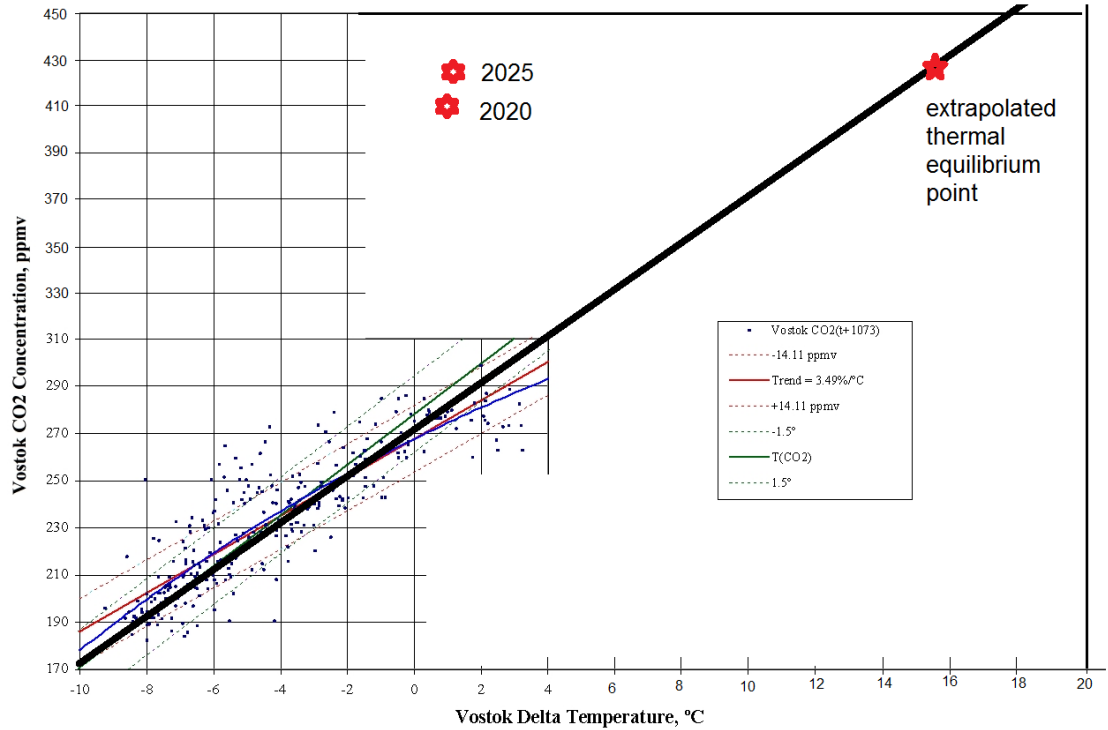

Appendix Figure 3: Correlation of the atmospheres $CO_2$ level and its temperature using the same data set as presented in Appendix Figure 1, showing here also the extrapolation of these two variables. Because of the emission of $CO_2$ as related to human activities the present CO2 levels, as shown for 2020 and 2025, are very high. Because of the known relationship between the increase of the global surface temperature of the atmosphere and the (Vostok Delat Temperature) and the atmosphere's $CO_2$ level, the extrapolated $CO_2$ level in ~ 2025 (~ 425 ppm), results in a estimated temperature response of approximately $\Delta T$ = 16 °C. Figure adapted from (Glassman, 2009), using the same VOSTOK $CO_2$ and temperature data as presented in (Petit et al., 1999). The extrapolated thermal equilibrium points are drawn from the $CO_2$ levels of 2020 and 2025. The thermal equilibrium points consider the historic relationship between the $CO_2$ level and the global warming. The data was obtained from the analysis of Earth's climate history obtained by hollow drillings of the Antarctic's ice sheet.

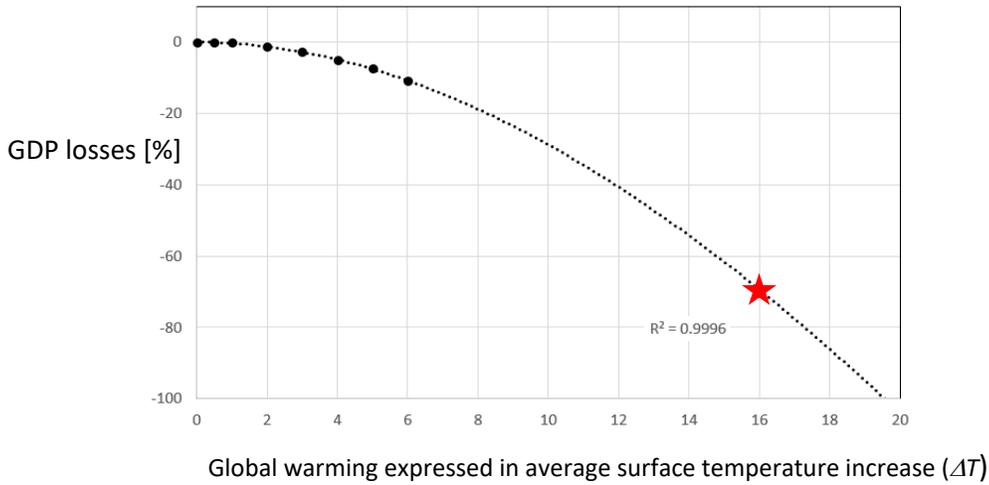

Appendix Figure 4: Estimated economic losses of the global gross domestic product (GDP) as a function of the average temperature increase of the Earth's surface, as related to the $CO_2$ emissions and global warming, with extrapolated data for the thermal equilibrium temperature increase of ΔT = 16 °C, as related to the extrapolated $CO_2$ level of 2025 (Appendix Figure 3). Figure adapted by extrapolation from (Stern, 2008).

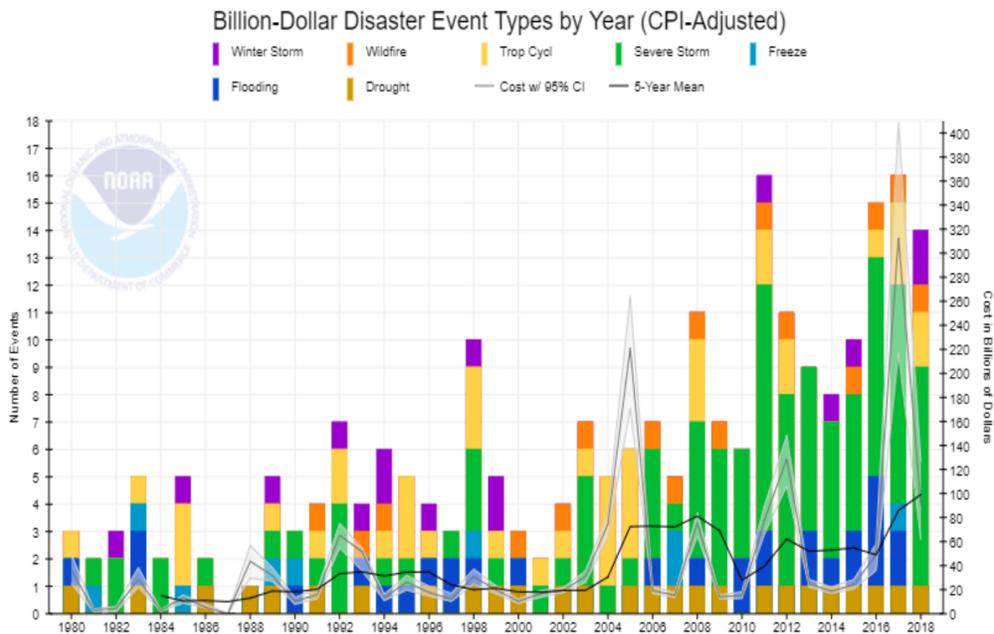

Appendix Figure 5: Tendency of the exponential increase of the frequency of billion disaster events, principally severe storms, as a function of time, all related to the change of the global climate because of the increased heat energy level in the atmosphere, because of the radiative forcing effect as principally related to the $CO_2$ level in the atmosphere.

Figure obtained from (NOAA, 2019).

## Appendix 4 - Explanation of related concepts and effects

**Thermal insulation effect of $CO_2$:** The known effect of the thermal insulation of $CO_2$ can be shown by a simple physics experiment using an infrared camera for the registration of thermal radiation of a human body. Filling a birthday air balloon with $CO_2$ and holding it between the camera and a person shows the balloon as a black spot, which is not the case if the balloon is filled with air.

**Radiative forcing and its energy:** Natural radiative forcing provides the Earth's necessary energy for being a habitable planet. While the radiative forcing of anthropogenic origin, as principally related to the $CO_2$ emissions, is only a small part of the natural radiative forcing, it adds large quantities of energy to the Earth's atmosphere and oceans. E.g. for 0.5 W/m$^2$ additive radiative forcing, the extra energy retained by the atmosphere corresponds to an equivalent heating with the energy of 400,000 Hiroshima nuclear bomb explosions per day (CLARK et al.), because of the Earth's large surface. This energy has to be comported by the Earth's atmosphere and hydrosphere. The present radiative forcing is with the value of circa 1 W/m$^2$ twice as high.

**Global warming:** The effect of the increase of the global average temperature of the atmosphere measured at sea level, is observed by temperature measurements, but is also simulated by a mathematical model which simulate the thermodynamic and behavior of the Earth's atmosphere at different heights and also includes the simulation of the Earth's hydrosphere. This modelling presents such a high sensitivity that temporal diming effects because of gases emitted from Vulcan eruptions can be visualized. The present global warming cannot be explained by other effects if in these simulations the warming effect as a function of the increased $CO_2$ level is taken out.

**Milankovich solar irradiation effect:** Milankovich discovered that the earth orbit and its declination is subjected to small modifications in a cycles of several thousand years. These changes resulted in the feeble increase or decrease of the absorbed solar irradiations by the atmosphere as a function of time, which resulted by hypothesis in the Earth's ice ages and warm periods as to see from Appendix Figure 2. Because of the preset declination earth is advancing to the next ice age. Meanwhile, the global warming related to the increased $CO_2$ level, compensates this cooling effect by a warming which is, however, approximately five order of magnitudes higher as the present Milankovich cooling.

**Climate history by south-pole ice core extraction:** Similar as the age and growth cycles of a three can be identified by the aging rings at a cut of its trunk, the climate history of the Earth can be accessed by the analysis of the ice layers of the south-pole, as built up by snow fall within several thousand years. In these layers air and water molecules are conserved as samples of the climate's history. As calibrated with present measurement of the heavy water isotope, warmer years lead to an increased evaporation rate of this heavy water isotope, and colder years lead to a decreased evaporation rate. By calibration an indirect measurement of the Earth's average temperature can be determined for a long term history of several thousand years (Appendix Figure 2). A measurement of the $CO_2$ level in the conserved air is also be calibrated with data from the present snow layers and $CO_2$ in the atmosphere resulting, therefore, in the indirect measurement of the $CO_2$ level.

**Difficulty to predict future extreme climate events:** Even so that the models which simulate climate change improved considerably over the past; they only can simulate the complex processes as related to the Earth's atmospheric and ocean circulations to a limited extend, because of the many interaction effects between the atmosphere, hydrosphere and biosphere. Therefore, many correlation effects were only discovered recently, and it is probable that many other effects will only be discovered in the future. As a result it is difficult to predict exactly the time when the severe outcomes of climate change will follow.

Examples of such effects are several positive feedback effects (Turner et al., 2020), tipping point effects, and domino effects (Klose et al., 2020), where a critical threshold of a first tipping element triggers several other tipping elements. A tipping point defines the transition of two stable states, which no point of regression. Because of the interactions between these elements a modelling which integrate all these elements and its nonlinear relationships is highly challenging, which makes the outcome of extreme events related to the future climate change highly unpredictable. While a modeling can lead to a most probable outcome, e.g., that of seawater rise, large model uncertainties lead to the increase of the uncertainty range especially as related to the time when such events will happen.

**Difficulty to mitigate climate change**: The problem of climate change is that the generated heat is produced as a function of two cumulative effects. The first is the cumulative accretion of $CO_2$ in the atmosphere, which generates the radiative forcing effect that results in the cumulative generation of heat absorbed by the Earth's hydrosphere and atmosphere as the incoming energy of short wave radiation is higher than the outgoing irradiation of long wave radiation (Shakoor et al., 2020). Assuming humankind would be very capable and stop the present $CO_2$ emission completely now, then the cumulative heat creation is still not solved, because of the invariant high value of the $CO_2$ in the atmosphere, which must be reduced, e.g., by reforestation measures. This means that the Earth's remaining ecosystem must be capable to convert the already emitted $CO_2$ in oxygen and carbon in order to avoid the further radiative forcing. Such a conversion is necessary as the already accumulated $CO_2$ in the atmosphere, as emitted since the industrial revolution, must first be converted by the Earth's vegetation to enable the reduction of the radiative forcing effect. This means, we must address not only the reduction of the $CO_2$ emissions by a 100% renewable energy generation, but we also must consider reforestation actions to reduce the present high level of $CO_2$ in the atmosphere to attain most swiftly pre-industrial levels. However, because of the two cumulative functions, these two actions do still not address the accumulated heat in the atmosphere.

**Climate change mitigation with the lowest risk**: Geoengineering could be a solution to solve this problem temporarily until the time when adequate measures can reduce the atmosphere's $CO_2$ content (Lutsko et al., 2020). However, the related unknown side effects, such as over- or undercooling present another threat. Therefore, and in order to avoid costs and risks related to (i) geoengineering and (ii) further climate change outcomes, we should as quick as possible transform the present energy generation and use, for a complete support with renewable power plants.

Reforestation actions in an adequate scale, and interspersed distancing, to avoid the proliferation of fire, also present low cost of US $ (5…50) / ton $CO_2$ capture and can provide additional income from the produced wood (Fawzy et al., 2020). As discussed by the Fawzy and Coworkers, the coverage of only 1% of the Earth's surface (500 million hectares), result in an additional $CO_2$ capture between 0.5 and 3.6 billion metrics tons $CO_2$ per year. The upper value corresponds to 9.5 % of the present annual emissions of $CO_2$. As a result large scale reforestation can reduce the present reduction of $CO_2$ and radiative forcing to preindustrial level. Presently, a small part of the reforestation action, saying the part which costs less than US $ 12 per captured ton of $CO_2$, can be financed by carbon taxes at an equivalent value, as this value refers to the present carbon taxes, as calculated from the table in (Wikipedia, 2021a), citing (Statistics, 2008), for different types of thermal power plants. Therefore, carbon taxes should be increased, and rigorously should also cover the present and expected damages from climate change.

**Understand exponential growth:** Exponential growth has the typical appearance of an inverted knee curve, at which growth starts at very low rates, and because of the continuous exponential increase growth rates become extremely large seaming to provide an almost vertical growth of the related effect after some time. In such a configuration it is obvious that time lags in the exponential growth, as eventually might happen for the cumulative installation of wind power, can limit the total growth as a function of time considerably. Therefore, adequate planning measures have to be implemented during the whole growth period, and the scale of these planning measures should be oriented on the exponential growth ability as here specified for the two presented VRE technologies. Additionally, the power grid must adapt adequately, especially for the high future VRE fractions, providing the necessary (i) storage, (ii) emulated storage, and (iii) grid stability measures as discussed in sections 2.1 to 2.10 in (Kratzenberg, 2021).

# Appendix 5 - A short picture of the primary energy generation with wind and PV power plants

By non-constrained exponential growth of VRE power plants we expect that in 2025 a total of 33,000 TWh/year will be generated by renewable power plants (Fig. 6). In this scenario the major part (~15,000 TWh/year) should be generated by photovoltaic power plants, the second major part (~ 11,700 TWh/year) by wind power plants, and the smallest part (~ 6,300 TWh/year) by hydropower plants (Fig.5). For such an expansion wind and PV technologies will be able to generate an MWh for a cost of less than US$10 (Fig. 8). Energy prices reduces furthermore for the 2030 scenario where 75,500 TWh should be generated by PV power plants, 31,100 TWh by wind power plants and 6500 TWh by hydropower plants, providing generation costs far below US$10 (Fig. 8). Such a low-cost, together with appropriate carbon taxes, enables to substitute many processes of primary energy use in an economic way with the use of renewable energy. E.g. the so called hydrogen economy (Kovač et al., 2021), will progress and expand, where hydrogen is used as fuel in hydrogen vehicles, for the heat generation, as energy storage, and for long distance transport in pipelines. Foundries and other industrial processes which generate the necessary heat by the combustion of fossil fuels can be converted to the combustion of hydrogen, as produced with the excess electrical energy in the power grid as generated by the VRE power plants. In vehicles, hydrogen fuel can be used for internal combustion engines (ICE) or the generation of electric energy via fuel cells. The somewhat lower efficiency as in comparison to electric cars based on Lithium-ion batteries will not be of importance because of the low generation cost of electric energy. Hydrogen ($H_2$), as generated by electrolysis, and $CO_2$ residuals can be used for the conversion of these two gases in natural gas ($CH_4$) by a so called methanation reaction, such as the well-known, and already used, Sabatier process. In such processes the necessary $CO_2$ can be captured from the exhaustion of thermal power plants. Furthermore, these thermal power plants can be operated with the synthetically generated methane, or natural gas (SNG) providing, therefore, a closed circle, which does not need the combustion of fossil fuel.

Optimized methanation plants are expected to provide electrical energy based on SNG at 40 Euro/MWh (Guilera et al., 2018), a similar value as compared to the present generation cost with thermal power plants based on natural gas (NG). Therefore, a cost effective provision of this synthetic fuel can be afforded usable in a large spectrum of different transport vehicles. Presently methanol is already used as rocket fuel in advanced SpaceX raptor engines, presenting the highest trust to weight ratio if compared to further engines ((Wikipedia, 2021d) citing further references).

The commercial aviation industry has to accelerate its development plans for the introduction of airplanes fueled with natural gas ($CH_4$) or hydrogen, which is presently predicted only for 2035 (Airbus, 2021) or beyond (Paur, 2012). Otherwise they may lose passenger shares to new flight concepts, presenting extremely streamlined aviation fuselages being, therefore, highly fuel efficient (MenkorAviation, 2021). As based on ICE this propeller driven aero plane type is able to fly with SNG in intercontinental flights. For intracontinental flights this vehicle can also operate with electric motors and propellers, where its energy is stored in lithium-ion batteries. With its commercial introduction predicted in 2025, this airplane presents an 8.6 times higher fuel economy, a 6.4 times lower operation cost, and a more than two times larger flight range, compared to a similar light passenger jets. Its flight speed of 724 km/h, is not losing much in comparison to conventional passenger jets.

## Appendix 6 – Further figures

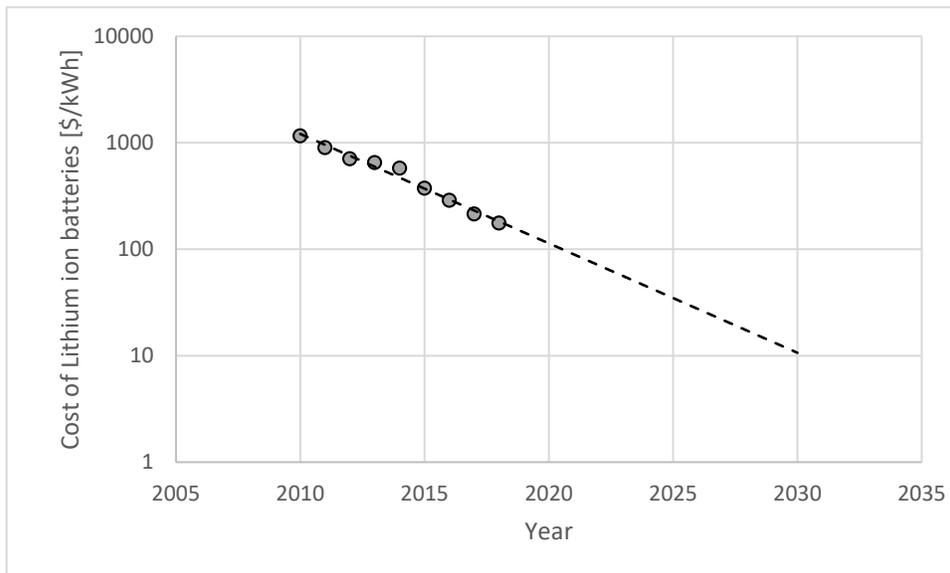

Appendix Figure 6: Historic and future cost estimation of the exponential cost decrease of Lithium-ion batteries. Figure adapted from (Martínez, 2020), citing (BloombergNEF, 2019).